\documentclass[conference]{IEEEtran}

\IEEEoverridecommandlockouts

\usepackage{amsmath,amssymb,amsfonts}
\usepackage{algorithmic}

\usepackage{graphicx}
\usepackage{textcomp}
\usepackage{xcolor}
\usepackage{colortbl}

\definecolor{causered}{rgb}{0.82,0.38,0.44}
\definecolor{effectblue}{rgb}{0.5,0.63,1}

\usepackage{cite}
\usepackage{hyperref}
\hypersetup{pdfborder=0 0 0}
\usepackage{cleveref}

\def\BibTeX{{\rm B\kern-.05em{\sc i\kern-.025em b}\kern-.08em
    T\kern-.1667em\lower.7ex\hbox{E}\kern-.125emX}}

\begin{document}

\title{CiRA: An Open-Source Python Package for Automated Generation of Test Case Descriptions from Natural Language Requirements\thanks{This work was supported by the KKS foundation through the S.E.R.T. Research Profile project at Blekinge Institute of Technology. The version of record of this article, first published in the \emph{Tenth International Workshop on Artificial Intelligence and Requirements Engineering} (co-located with \emph{31st IEEE International Requirements Engineering 2023 conference}), is available online at Publisher’s website: \url{https://doi.org/10.1109/REW57809.2023.00019}}
}

\author{
    \IEEEauthorblockN{
        1\textsuperscript{st} Julian Frattini \\
        3\textsuperscript{rd} Andreas Bauer \\
        }
    \IEEEauthorblockA{\textit{Blekinge Institute of Technology}\\
        Karlskrona, Sweden \\
        \{firstname\}.\{lastname\}@bth.se}
    \and
    \IEEEauthorblockN{
        2\textsuperscript{nd} Jannik Fischbach}
    \IEEEauthorblockA{\textit{Netlight Consulting GmbH and fortiss GmbH}\\
        Munich, Germany \\
        jannik.fischbach@netlight.com}
}

\maketitle

\begin{abstract}
    Deriving acceptance tests from high-level, natural language requirements that achieve full coverage is a major manual challenge at the interface between requirements engineering and testing. Conditional requirements (e.g., ``If A or B then C.'') imply causal relationships which---when extracted---allow to generate these acceptance tests automatically. This paper presents a tool from the CiRA (Causality In Requirements Artifacts) initiative, which automatically processes conditional natural language requirements and generates a minimal set of test case descriptions achieving full coverage. We evaluate the tool on a publicly available data set of 61 requirements from the requirements specification of the German Corona-Warn-App. The tool infers the correct test variables in 84.5\% and correct variable configurations in 92.3\% of all cases, which corroborates the feasibility of our approach.
\end{abstract}

\begin{IEEEkeywords}
    requirements engineering, natural language processing, acceptance test, test case description, BERT
\end{IEEEkeywords}

\section{Introduction}
\label{sec:intro}

A key challenge of high-level acceptance testing---where the behavior of a system is compared to the behavior specified by the requirements---is generating a set of test cases that fully cover a requirement~\cite{whalen2006coverage}. A requirement is fully covered when the test cases associated with it assert all configurations of input and output variables implied by the requirement. 

Conditional requirements (e.g., ``\textbf{When} the red button is pushed \textbf{or} the power fails \textbf{then} the system shuts down.'') are often used to specify functional requirements~\cite{frattini2023causality} and make the causal relationship between input and expected output explicit~\cite{fischbach2021practitioners}. This causal relationship can be used to construct a minimal set of test cases fully covering the requirement~\cite{fischbach2020towards}.

Since these conditional requirements can be detected with a reasonable accuracy~\cite{fischbach2021automatic}, approaches to automatically extract the causal relationship from the natural language (NL) requirement have been developed~\cite{fischbach2023automatic}. We present a tool to automatically generate a minimal set of test case descriptions from an NL requirement, ensuring its full coverage. We disclose all source code and demonstrate the tool's applicability by generating test case descriptions for the Corona-Warn-App.

\section{Tool Introduction}
\label{sec:tool}

% use case and pipeline
The goal of the CiRA tool is to generate a minimal but fully covering set of test case descriptions from a single-sentence, natural language requirement specification implying a causal relationship. To achieve this goal, the tool needs to (1) identify whether the sentence contains a causal relationship, (2) assign each word in the sentence to its role in the causal relationship, (3) transform the NL sentence into a cause-effect graph (CEG), and finally (4) derive a minimal set of test case descriptions from the CEG. We illustrate these four steps of the tool's pipeline with the example sentence from the introduction.

\textbf{Step 1: Identifying causal relationships.} First, the CiRA tool performs a binary classification to categorize the NL requirements specification as either \texttt{causal} or \texttt{non-causal}~\cite{fischbach2021automatic}. Only causal sentences can be processed further. In this example, the \textit{CiRA classifier} module categorizes the sentence as \texttt{causal} with 98\% confidence.

\begin{figure}[h]
    \centering
    \includegraphics[width=0.48\textwidth]{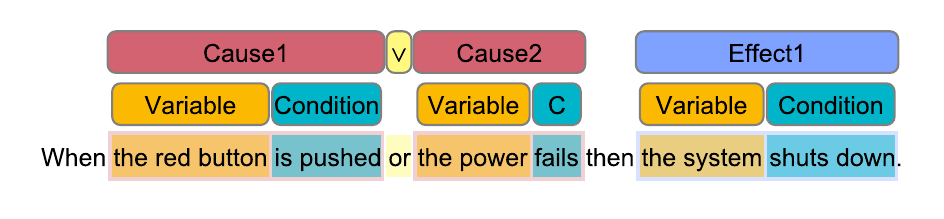}
    \vspace{-0.5cm}
    \caption{Labeled requirements specification}
    \label{fig:labeling}
    \vspace{-0.2cm}
\end{figure}

\textbf{Step 2: Assigning roles to words.} Secondly, the CiRA tool performs a multi-labeling task to assign to each word of the sentence its respective role in the causal relationship. There are two levels of labels: (1) \textit{event labels}, which determine to which event (cause or effect) a word belongs, or \textit{junctors}, which determine how events are related, and (2) \textit{sub-labels}, which determine the part of an event (variable or condition) a word belongs to. \Cref{fig:labeling} visualizes how the \textit{CiRA Labeler} annotates the example sentence: it identifies three events, and the two cause-events are related with a disjunction (labeled $\lor$). In each event, its variable is distinguished from its condition.

\begin{figure}[h]
    \centering
    \includegraphics[width=0.48\textwidth]{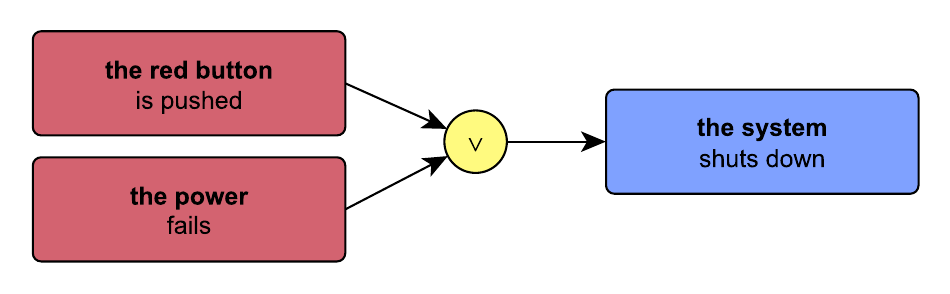}
    \vspace{-0.5cm}
    \caption{Generated cause-effect graph}
    \label{fig:ceg}
    \vspace{-0.2cm}
\end{figure}

\textbf{Step 3: Constructing a graph.} Thirdly, the CiRA tool constructs a Cause-Effect-Graph (CEG) from the labeled sentence. The \textit{CiRA Graph Constructor} generates one node for each event and connects events with directed edges according to their relationship. \Cref{fig:ceg} shows the translation of the labeled sentence into a CEG.

\textbf{Step 4: Deriving test cases.} Finally, the CiRA tool derives a minimal set of test case descriptions from the CEG, which fully covers the initial requirements specification. For this, the \textit{CiRA Test Case Generator} considers all events to be Boolean (i.e., they can be evaluated to true or false) and determines all configurations of the cause-events necessary to evaluate the effect-events to both true and false. \Cref{tab:testsuite} shows the test suite generated for the example sentence. To evaluate the effect-event to be false, both cause-events need to be false as well (TC3). To evaluate the effect-event to be true, either one of the cause-events must be true (TC1 and TC2) given their connection via a disjunction (i.e., logical or). A fourth test case where both cause-events are true is unnecessary, as TC1 and TC2 assert that either of the two events can cause the effect-event to be evaluated as true.

\begin{table}[h]
    \centering
    \caption{Generated test suite of test case descriptions}
    \label{tab:testsuite}
    \begin{tabular}{|l|ll|l|} \hline
        \textbf{ID} & \cellcolor{causered}\textbf{the red button} & \cellcolor{causered}\textbf{the power} & \cellcolor{effectblue}\textbf{the system} \\ \hline
        TC1 & is pushed & \textit{not} fails & shuts down \\
        TC2 & \textbf{not} is pushed & fails & shuts down \\
        TC3 & \textbf{not} is pushed & \textbf{not} fails & \textbf{not} shuts down \\ \hline
    \end{tabular}
\end{table}

\begin{figure*}[h]
    \centering
    \includegraphics[width=0.9\textwidth]{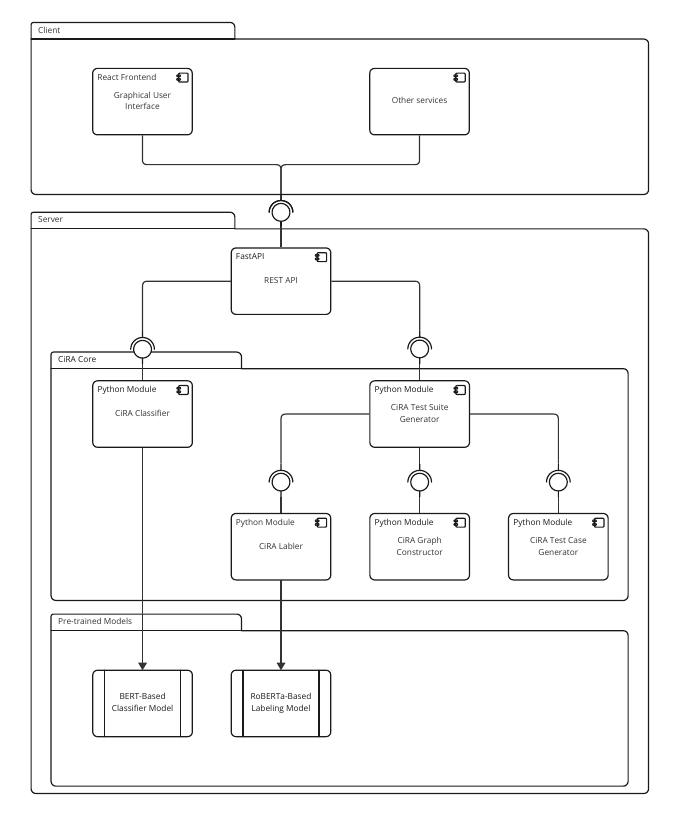}
    \caption{Architecture of the CiRA Tool}
    \label{fig:cira-architecture}
\end{figure*}

\begin{figure*}
    \centering
    \includegraphics[width=\textwidth]{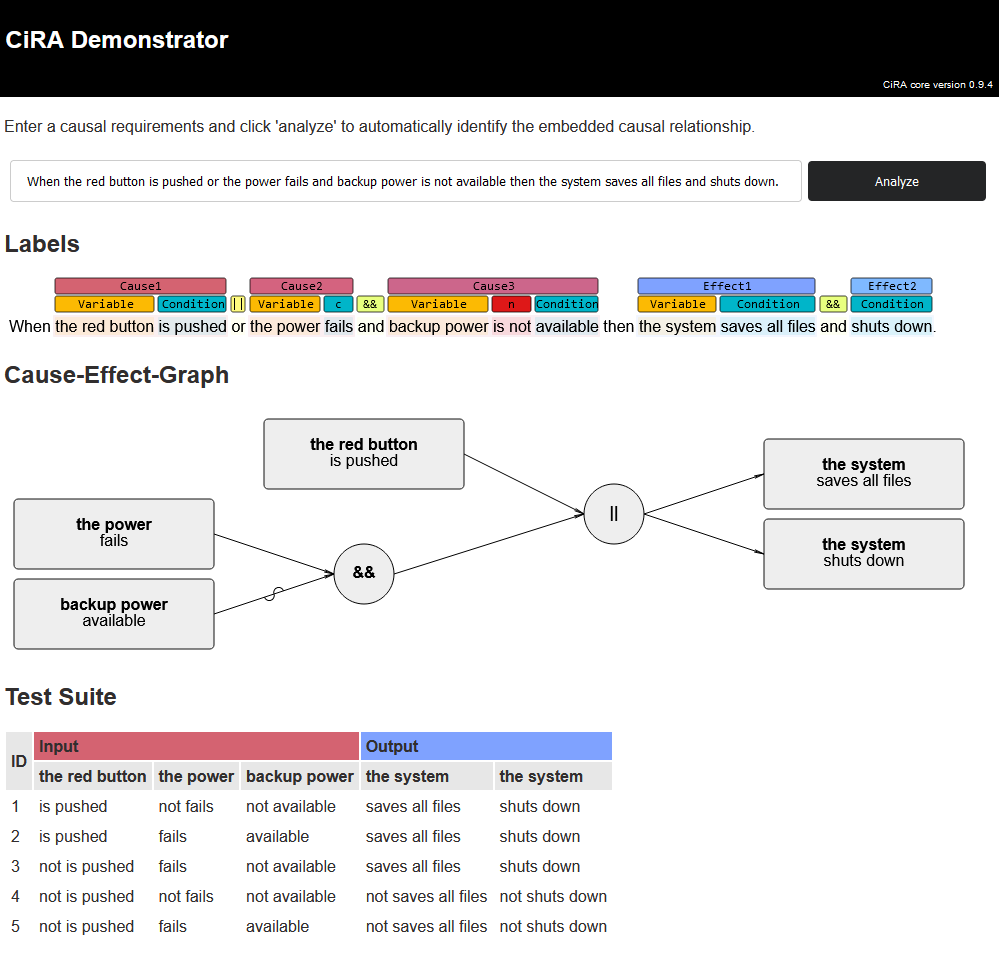}
    \caption{User Interface for the CiRA tool}
    \label{fig:cira-ui}
\end{figure*}

% architecture
In previous work~\cite{fischbach2021automatic,fischbach2023automatic}, we developed a prototype of this tool for evaluation purposes. This paper presents a maintainable, usable, and open-access evolution of that prototype. The CiRA tool---visualized in \Cref{fig:cira-architecture}---consists of the CiRA core\footnote{Source code at \url{https://zenodo.org/badge/latestdoi/456568427}}, which offers the functionality of the described pipeline. The classification and the labeling task utilize pre-trained, fine-tuned machine learning models based on BERT and RoBERTa~\cite{frattini2023causality,fischbach2023automatic}. A REST API provides an interface to the functionality of the CiRA tool, which is used by a dedicated graphical user interface (GUI)\footnote{Source code at \url{https://zenodo.org/badge/latestdoi/571614601}} for human interaction\footnote{Online demo at \url{www.cira.bth.se/demo}}. \Cref{fig:cira-ui} shows the full GUI with another, more complex example.

\section{Demonstration}
\label{sec:demo}

% data set
To demonstrate the usability and performance of the CiRA tool, we evaluate its capability on an unseen set of natural language requirements\footnote{Data and code at \url{https://zenodo.org/badge/latestdoi/650042294}}. Similar to the previous version of the tool~\cite{fischbach2021cira}, we chose the requirements of the German Corona Warn App\footnote{Requirements specification at \url{https://github.com/corona-warn-app/cwa-documentation/blob/main/scoping_document.md}} as a target system due to its recency, openness, and realism. The system contains ten epics, 32 user stories connected to these epics, and in total, 61 acceptance criteria. 

% method
The subject of the evaluation is the 72 natural language sentences constituting the acceptance criteria. We manually classified each sentence as either \textit{causal} (N=26) or \textit{non-causal} (N=46) and created an expected set of test cases for each of the causal sentences. Then, we executed both the CiRA classifier and the CiRA Test Suite Generator and compared the automatically generated results with the manual ones. We evaluate the classifier using the macro F1-score~\cite{fischbach2021cira} and the test suite generator regarding its ability to infer correct events from the causal sentence and its ability to determine the correct test value configurations.

% results
The CiRA Classifier achieves a macro f1-score of 79.26\%. The CiRA test suite generator infers the correct test variables in 84.5\% and the correct test value configurations in 92.3\% of all cases. The results show that CiRA can automate a large part of manual test case generation but struggles with implicitly causal or ill-structured sentences.

%\section{Conclusion and Future Work}
%\label{sec:conclusion}

% future work
While the CiRA tool was mainly designed for test case generation, the intermediate pipeline steps allow using it in further use cases. Future research could include completeness assessment (i.e., assessing the output of the labeling to evaluate whether all components of a causal relationship (variables and conditions) are contained in the sentence) and dependency detection (i.e., comparing the CEGs of multiple requirements to identify dependencies). Finally, connecting the pipeline to another tool that converts the test case descriptions into executable test cases will fully automate the acceptance testing process.

\bibliographystyle{IEEEtran}
\bibliography{material/references}

\end{document}